\documentclass[usenatbib]{mn2e}

\title[Polarisation of prestellar cores]{SCUBA polarisation observations 
of the magnetic fields in the prestellar cores L1498 and L1517B}

\author[Kirk, Ward-Thompson \& Crutcher]
{J. M. Kirk$^{1,2}$, D. Ward-Thompson$^{1,3}$, R. M. Crutcher$^2$ \\
$^1$School of Physics and Astronomy, Cardiff University,
5 The Parade, Cardiff, CF24 3YB \\
$^2$Department of Astronomy, University of Illinois, 1002 West Green Street,
Urbana, Illinois, IL61801, USA \\
$^3$Observatoire de Bordeaux, 2 rue de l'Observatoire, 33270 Floirac,
France}

\date{Accepted 2006 March ?; received 2006 March 24; in original form
2006 January 17.}

\pagerange{000--000}

\begin{document}

\label{firstpage}

\maketitle
\begin{abstract}
We have mapped linearly polarized dust emission from the prestellar 
cores L1498 and L1517B with the James Clerk Maxwell Telescope (JCMT) using 
the Submillimetre Common User Bolometer Array (SCUBA) and its polarimeter 
(SCUBAPOL) at a wavelength of 850~$\mu$m. We use these measurements to
determine the plane-of-sky magnetic field orientation in the cores.
In L1498 we see a magnetic field 
across the peak of the core that lies at an offset of 
$\sim$19$^\circ$ $\pm$ 12$^\circ$ to the short axis of the core.
This is similar to the offsets seen in previous observations of prestellar
cores. To the southeast of the peak, in the filamentary
tail of the core, we see that the 
magnetic field has rotated to lie almost parallel to the long axis of the
filament. We hypothesise that the field in the core may have
decoupled from the field in the filament that connects the core
to the rest of the cloud. We use the Chandrasekhar-Fermi (CF) method to
measure the plane-of-sky field strength in the core of
L1498 to be $\sim$10 $\pm$ 7 $\mu$G.

In L1517B we see a more gradual turn in the field direction 
from the northern part of the core to 
the south. This appears to follow a twist in the filament in
which the core is buried, with the field staying at a roughly 
constant $\sim$25$^\circ$ $\pm$ 6$^\circ$ offset to the short axis of the
filament, consistent with previous observations of prestellar cores.
Hence these two clouds in an apparently similar
evolutionary state, that exhibit similar masses,
morphologies and densities, have very different magnetic field configurations.
We again use the CF method and calculate the magnetic field
strength in L1517B to be $\sim$30 $\pm$ 10 $\mu$G.
Both cores appear to be roughly virialised. Comparison with our previous
work on somewhat denser cores shows that, for the denser cores, thermal and 
non-thermal (including magnetic) support are approximately equal, while 
for the lower density cores studied here, thermal support dominates.
\end{abstract}
\begin{keywords}
ISM: clouds -- ISM: individual (L1498, L1517B) -- ISM: 
magnetic fields -- polarization -- stars:formation
\end{keywords}

\section{Introduction}\label{intro}

The exact manner in which stars form is still a matter of debate. One 
important aspect of this debate centres around the morphology and 
magnitude of the magnetic field within molecular clouds and the 
magnitude of its effect on the star formation process.
Observations of the magnetic field before protostellar collapse are
necessary to address this question, ideally in a sample of molecular clouds
that do not contain protostars but are on the verge of protostellar collapse.

Myers and co-workers developed a catalogue 
of dense molecular cores within dark clouds, which they observed in CO 
and later in ammonia \citep[see:][and references therein]{bm89}. 
\citet{beichman86} cross-referenced the Myers ammonia catalogue 
with the IRAS point source catalogue and identified starless cores 
as those that had not yet formed an embedded protostar. 
\cite{wsha94} observed the submillimetre continuum emission from 
starless cores and identified a 
subset of starless cores called prestellar (formerly 
pre-protostellar) cores that are thought to be about to form 
a protostar. 

\citet[][hereafter KWA]{kirk05} observed a sample of 52 candidate 
prestellar cores with 
the Submillimetre Common User Bolometer Array (SCUBA) camera on the 
James Clerk Maxwell Telescope (JCMT). Twenty-nine of the cores were 
detected and were separated into `bright' and `intermediate' groups 
based on their 850-$\mu$m peak flux densities. The bright cores were 
seen to be more centrally condensed. Following the method of 
\citet{jessop01}, KWA showed that the estimated lifetimes of the 
two groups were consistent with them being part of an evolutionary sequence.  

Ward-Thompson et al. (2000)
presented the first observations of the magnetic field 
geometry from prestellar cores when they mapped L183, L1544 and L43 
with the SCUBA polarimeter. Their maps showed relatively smooth and 
uniform magnetic fields over the central core regions that were at an 
angle of approximately $30\degr$ to the projected minor axis of the cores. 
\citet{cnwk04} used the \citet{dwt00} data to estimate the strength 
of the magnetic field in the plane of the sky using an effect first 
suggested by \citet[hereafter CF]{cf53}. 

CF postulated that when 
the magnetic field is frozen to the matter, turbulent motions will cause 
local perturbations in a uniform magnetic field. These perturbations to 
the magnetic field should manifest themselves as a random scatter 
superimposed on an otherwise regularly oriented magnetic field patern. 
If the magnitude of the random component of the dispersion, 
$\delta \phi$, and the turbulent motion, $\delta V_{NT}$, 
are known then CF showed 
that it is possible to infer the strength of the uniform plane-of-sky
field $B_{pos}$ in a medium of density $\rho$, such that:

\begin{equation}
B_{pos} = Q\sqrt{4\pi\rho}\, \frac{\delta V_{NT}}{\delta \phi} \approx 9.3 
\sqrt{n(H_{2})} \frac{\Delta V_{NT}}{\delta\phi} \mu G,
\end{equation}  

\noindent
where $\rho = m n(H_{2})$, $m$ is the mean particle mass, $n(H_2)$ is
the volume number density
and $\Delta V_{NT} = \delta V_{NT} \sqrt{8\ln 2}$ is the 
full-width at half maximum
(FWHM) of the non-thermal, or turbulent, linewidth.
$Q$ is a calibration factor, 
fixed by comparison with simulations, for which we follow
\citet{cnwk04} in adopting a value of 0.5.

A crucial parameter in the study of magnetically influenced star formation 
is $M/\Phi$, which is the ratio of the mass to the magnetic flux and is 
normally expressed as $\lambda$, the ratio of the observed mass-to-flux ratio 
$(M/\Phi)_{obs}$ to the critical mass-to-flux ratio
$(M/\Phi)_{crit}$. Above the critical mass-to-flux ratio
the magnetic field cannot 
support the observed mass against gravitational collapse. 

Inferring $\lambda$ 
from observations is possible if the column density $N$ and the magnetic 
field strength $B$ can be determined such that:
\begin{equation}
\lambda = \frac{(M/\Phi)_{obs}}{(M/\Phi)_{crit}} = 
\frac{(mNA/BA)}{(1/2\pi\sqrt{G})} = 7.6\times10^{-21} \frac{N(H_{2})}{B},
\end{equation}
where $m = 2.8m_{H}$, allowing for 10\% He by number, with $N(H_{2})$ in 
cm$^{-2}$ and B in $\mu$G. 

\cite{cnwk04} estimated that the three cores they 
analysed had $B_{pos} \sim 80-160 \mu$G and $\lambda \sim 1.9-2.6$. When 
corrected downwards by a statistical
factor of three for geometrical bias this showed 
that the magnetic field strength was such as to render each
core slightly magnetically subcritical. 

However, both Ward-Thompson et al. (2000) and
\cite{cnwk04} pointed out that the offset between the core
short axis and the magnetic field direction was not consistent with
magnetically-regulated star formation models \citep[e.g.][and references
therein]{cm98}.
These models predict preferential collapse along the short axis of
the core, yielding a field direction parallel to the short axis.
This was explained as a projection effect in triaxial cores
(Basu 2000; Ciolek \& Basu 2004), and it was claimed
that the data were consistent with the models.

All three of the prestellar cores
previously mapped with the SCUBA polarimeter belong 
to the group of bright cores in the KWA survey. In this paper we 
present data for
two more cores, L1498 and L1517B, which are somewhat less luminous.
L1517B is a bright core and L1498 is in the intermediate group,
although both lie near the borderline, and they are in fact quite
similar. The goal is
to examine whether the magnetic field morphology
and/or strength change with 
submillimetre brightness, mass or evolutionary stage. 

We describe 
the observational parameters in the next Section and present our data in 
Section \ref{results}. In Section \ref{discuss} we apply the CF technique 
to the new data and 
calculate various physical parameters of the cores.

\section{Observations}\label{obs}

Submillimetre continuum observations at 850~$\mu$m were carried out 
using SCUBA \citep{scuba99} on the JCMT on
Mauna Kea, Hawaii, during the evenings of 1999 August 13--15 between 
the hours 01:30 and 09:30 HST (11:30--19:30 UT) and on the evenings of 2004 
February 2--3 between the hours 17:30 and 01:30 HST (03:30--11:30 UT). 

SCUBA was used with the SCUBAPOL \citep{greaves03}
polarimeter, which uses a rotating half-waveplate and fixed analyser. 
The waveplate is stepped through sixteen positions 
(each offset from the last by 22.5$\degr$) and a Nyquist-sampled image 
(using a 16-point jiggle pattern) is taken at each waveplate position
\citep{greaves03}. 
The observations were carried out whilst chopping the secondary 
mirror 120 arcsec in azimuth at 7 Hz and synchronously detecting the signal, 
thus rejecting sky emission. 

The integration time per point in the 
jiggle cycle was 1 second, in each of the left and right telescope 
beams of the dual-beam chop.
The total on-source integration time per complete cycle was 512 
seconds. Two sources were observed, L1498 and L1517B, and the entire
observing process described above 
was repeated 46 times for L1498 and 32 times for L1517B. 

The instrumental polarisation (IP) of each bolometer was measured on the 
planets Mars and Uranus. This was subtracted from the data before calculating 
the true source polarisation. The mean IP was found to be 0.93
$\pm$ 0.27\%. It was found necessary to offset the pointing between 
repeated observations so that holes left by the exclusion of bolometers 
with significantly higher than average noise statistics could be filled 
in. The submillimetre zenith opacity for atmospheric extinction removal 
was determined by comparison with the 1.3-mm sky opacity \citep{archibald02}. 
The average 850-$\mu$m zenith opacity was 0.28 during the 1999 observations
and 0.19 during the 2004 observations, 
corresponding to a mean zenith transmission of 76 and 83\%
respectively.  

\section{Results}\label{results}

Figures \ref{l1498} and \ref{l1517b} show maps of the prestellar cores 
L1498 and L1517B. The greyscale and
contours show the Stokes I (intensity) of the
dust emission. The absolute calibration of the intensity is not
maintained through polarimetry observations.
Therefore we simply show the contours as a fraction of
the peak intensity and refer to previous observations to obtain
the absolute intensity calibration and parameters derived therefrom.
For example,
KWA found peak flux densities of 120 and 170~mJy/beam and integrated
flux densities of 2.3 and 2.6~Jy respectively for L1498 and L1517B at
850~$\mu$m. We use these data below to calculate core
masses and densities.

We note that there is a
180$^\circ$ ambiguity in the field direction, as two anti-parallel fields
would generate the same polarisation. The correct mathematical term for
such a vector with a 180$^\circ$ ambiguity is a `half vector'.
Therefore, this is the term that we use.
The half vectors in Figures 1 \& 2 show the 
direction of the magnetic field in the plane of the sky. This is inferred 
from the assumption that the dust grains are magnetically aligned and that 
the magnetic field direction is orthogonal to the direction of polarisation 
\citep{dg51}. Therefore the polarisation half vectors have been rotated by 
90$^\circ$ to indicate the field direction. 

The half vectors in Figures 1 \& 2 are binned at 12-arcsec spacing 
to roughly match the JCMT beam FWHM of $\sim$14 arcsec at 850~$\mu$m. 
Only those half vectors with a signal-to-noise ratio $\geq$2~$\sigma$ are
plotted. This corresponds to a maximum position angle uncertainty of 
$\pm$14$^\circ$. For each core the highest detection was found to be just 
under 8$\sigma$ and the mean half vector signal-to-noise ratio
was $\sim$4--5~$\sigma$.

Note that, whenever we 
quote an error-bar on a polarisation half vector position
angle measurement, we are quoting the observational error rather than the
dispersion observed from one half vector to another. This error can be 
calculated for a single half vector from its signal-to-noise ratio, 
based on its stability
as a function of time throughout the observations. This can be cross-checked
using the known sensitivity of the instrument, the observing conditions and
the total integration time. The error-bar for a group of half vectors can then
be calculated from each half vector's individual error-bar.

\subsection{L1498}

\begin{figure}
\setlength{\unitlength}{1mm}
\noindent
\begin{picture}(80,65)
\put(0,0){\includegraphics{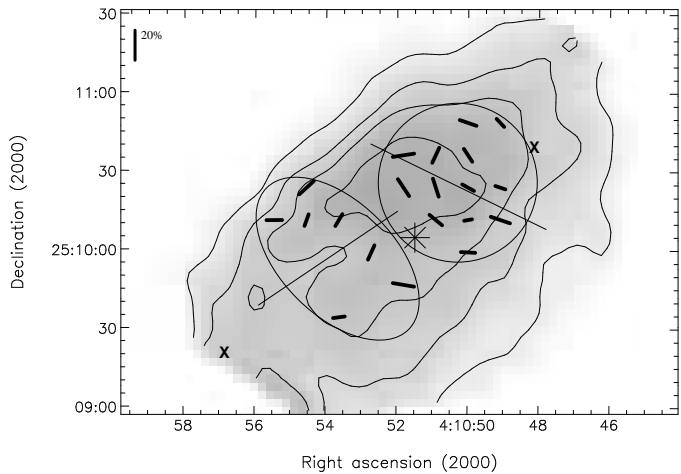}}
\end{picture}
\caption{The prestellar core L1498 seen in submillimetre continuum dust 
emission at a wavelength of 850~$\mu$m. The contours are at
40, 60, 80 \& 90\% of peak intensity. 
The polarisation half vectors have been rotated by 90$^\circ$
to show the magnetic field direction on the plane of the sky. 
The length of each half vector is proportional to the percentage polarisation.
The ovals show groups of half vectors with roughly similar orientations 
and the line through each oval shows the approximate mean weighted PA of 
each group. The crosses 
mark the centres of integrated CCS emission observed 
by \citet{lai2000}. The star marks the location of the 
CCS Zeeman observation undertaken by \citet{levin2001}.}
\label{l1498}
\end{figure}

Figure 1 shows an intensity map of L1498 at 850~$\mu$m as a 
greyscale with contours 
overlaid. Superposed on this are a series of polarisation half vectors which 
have been rotated by 90$^\circ$ to illustrate the magnetic field direction. 
The length of each half vector is proportional to the percentage polarisation 
observed. Looking first at the morphology of the core,
we see an elongated core extending northwest to southeast, with
a brighter peak at its northwestern end and a tail extending to the southeast.
KWA also mapped L1498 at 850~$\mu$m, but in photometric
mapping mode only, and observed a slightly more rounded morphology.

The intensity map 
presented in Figure~\ref{l1498} is deeper and larger in extent
than the previous 
map and shows that the chop direction chosen for the original map was 
almost parallel to the long axis of the core (the orientation of the core 
on these scales was not known at the time of the original observations). 
The chop function of the KWA data
(see their section 5.4) caused an underestimate
of the structure parallel to the major axis of the core and an offset in the 
apparent position angle of the core. It also led to an underestimate of 
the magnitude of the extended flux density. Correcting for the chop function
takes L1498 to being a borderline `bright' core, even more similar to
L1517B.

The map of L1498 in Figure~\ref{l1498} shows an elliptical core with a 
position angle (PA) for the long axis of the extended emission of 
$\sim$135$^\circ$ $\pm$ 5$^\circ$. 
The flux distribution is asymmetric along the long 
axis as mentioned above.
The far-infrared maps made by the Infrared Space Observatory (ISO) 
of this region show that the
L1498 core is at the northwestern end of a filament extending 
from the main cloud which lies to the southeast (Ward-Thompson,
Andr\'e \& Kirk 2002).

The surface of this filament appears to be being heated by the 
inter-stellar radiation field
(ISRF) and has a higher colour temperature 
than the interior of the core \citep{wak02}. The 
PA of the major axis of the core seen in our new data
is coincident with the PA of the filament
seen in the ISO maps (c.f. figure~7 of KWA).
The tail of emission seen in Figure~1
heading to the southeast of the core is the start of this filament.

However, there is an anti-correlation between the peak of the 850-$\mu$m
emission and the peak of integrated CCS and CS emission. The latter peaks 
towards the SE of the core \citep{taff04,lai2000}. Figure 8 of 
\citet{lai2000} shows what appears to be a CCS ring 
shape with a central depleted `hole'. The peaks of the CCS
emission are shown as crosses in Figure \ref{l1498}. 

We detect twenty polarisation half vectors across L1498 above the 2$\sigma$ 
cut-off. There appears to be a dramatic
change in field direction from the half vectors 
in the brighter northwestern part of the core to those in the southeastern 
part. The two groups are shown encircled on Figure \ref{l1498}.
In between there is a gap in the half vectors where presumably the two
different field directions compete and cancel the net polarisation. 

The thirteen half vectors on the core peak
have a weighted mean PA of 64$\degr$ $\pm$ 7$^\circ$, 
which is at an angle of $19\degr$ $\pm$ 12$^\circ$ 
to the PA of the short axis of the core. This alignment offset
of the magnetic field to the orientation of the core short axis
is similar to the $\sim$20--30$^\circ$ offset
found in previous observations of prestellar cores
\citep{dwt00}, as discussed in section 1 above.

The seven half vectors away from the core peak in the southeastern tail
have a weighted mean PA of 124$\degr$ $\pm$ 6$^\circ$
and are approximately parallel (11$^\circ$ $\pm$ 11$^\circ$ offset)
to the long axis of the core.
This rotation of the magnetic field through 
60$^\circ$ $\pm$ 11$^\circ$ is very marked and
is most unusual. It has not been observed in a prestellar 
core before \citep{dwt00}.
We note that the seven half vectors in the core's tail
are coincident with the `hole' in CCS emission discussed above,
and with a region of anomalous velocity seen in N$_2$H$^+$
(Tafalla et al. 2004).

However, we noted above that the tail to the southeast seen in Figure~1
is also seen in the ISO data to extend even further to the southeast,
as a filament linking up to the rest of the cloud (see figure~7 of KWA).
So it may be that the extended filament has a
different B-field direction from the dense core at its head
and that the two fields have essentially become decoupled. If that is the
case, then it seems
quite remarkable that the two fields have become so decoupled
as to lie at such a large apparent angle to one another.

\subsection{L1517B}

\begin{figure}
\setlength{\unitlength}{1mm}
\noindent
\begin{picture}(80,85)
\put(0,0){\includegraphics{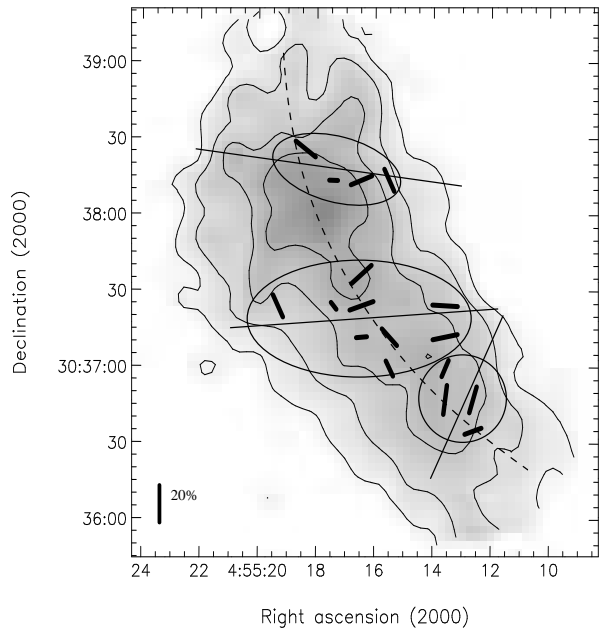}}
\end{picture}
\caption{The prestellar core L1517B in submillimetre dust emission at 
850~$\mu$m. The contours are at 40, 60, 80 \& 90\% of peak intensity.
The polarisation half vectors have again been rotated by 90$^\circ$ 
to show the plane-of-sky magnetic field direction. 
The ovals show groups of half vectors of similar orientation
and the lines through the ovals show the approximate mean weighted 
PA of those half vectors.
The dashed line shows the ridge of the larger filament
in which the core is embedded.}
\label{l1517b}
\end{figure}

The map of L1517B in Figure \ref{l1517b} shows a peak of
emission with a tail that extends to the southwest.
The composite ISO and SCUBA maps of KWA show 
that L1517B extends in a truncated filament northeast of
the main L1517 cloud. 
The peak of submillimetre intensity is near the `head' of 
this filament and the tail points back towards the main cloud. 
Hence it has a somewhat similar morphology and situation to L1498,
with a core near the head of a filament.

However, there is also some evidence in the intensity
map of Figure~2 that there is a slight extension of the filament
to the north of the peak, and that the filament turns slightly. 
Furthermore, the ISO data also show some evidence for a twist in the
filament just to the north of L1517B (see figure~7 of KWA).
Hence we see that the elongated filament in which the peak sits
turns slightly from its northern end to its southern part. This
is illustrated in Figure~2 by a curved dashed line which follows
the ridge of the filament.

We detected seventeen polarisation half vectors across L1517B above the 
2-$\sigma$ cut-off. They have a weighted mean PA of 
106$\degr$ $\pm$ 7$^\circ$. However, 
the morphology of the half vectors does not at first appear to 
be a simple uniform pattern. They were split into three groups
of similar orientation half vectors -- the northern four half vectors 
(PA $\sim$82$\degr$ $\pm$ 6$^\circ$), 
the middle nine half vectors (PA $\sim$94$\degr$ $\pm$ 6$^\circ$), and the 
southern four half vectors (PA $\sim$156$\degr$ $\pm$ 5$^\circ$) -- 
as shown in 
Figure~\ref{l1517b}. There appears to be a progression in mean PA from 
the north to the south, indicating a change in the mean field 
direction.

The PA of the long axis of the filament was measured at the centre of
each oval, and found to be 15$^\circ$ $\pm$ 5$^\circ$,
35$^\circ$ $\pm$ 5$^\circ$ and 50$^\circ$ $\pm$ 5$^\circ$ from
north to south respectively.
Therefore the angle between the field direction and the filament
elongation axis is 67$^\circ$ $\pm$ 11$^\circ$,
59$^\circ$ $\pm$ 11$^\circ$ and 75$^\circ$ $\pm$ 10$^\circ$ from
north to south respectively.

The mean of these is 65$^\circ$ $\pm$ 6$^\circ$, and hence the
mean of the offset from the short axis is
25$^\circ$ $\pm$ 6$^\circ$, consistent with that seen in the
core of L1498, and with previous observations of prestellar
cores \citep{dwt00}. So we see that the field is simply
following the twist in the filament and remaining at a roughly 
constant offset to it, as was seen before.
However, we note that the sign of the offset flips in the south.
This flip cannot be explained by the argument that this is simply a
projection effect in a triaxial core, as suggested by the proponents
of the magnetically regulated models (Basu 2000; Ciolek \& Basu 2004).

\section{Discussion}\label{discuss}

We can use our measurements not only to map the orientation of the
magnetic field, but also to estimate the strength of the field in the
plane of the sky, using the CF method, as discussed in section 1 above.
For this we need to measure the dispersion in the polarisation half vectors
(corrected for dispersion caused by observational errors), and to compare
this with the mass, density and non-thermal spectral linewidth in each core.

The densities of the cores were estimated by
assuming that they are triaxial ellipsoids,
with the line-of-sight axis being the mean of the other two axes. 
We use the continuum data of KWA to calculate the dust mass, and hence
the total mass of each core. The conversion from 
total mass to mass of $H_{2}$ is made using the standard solar abundance 
of H. These values were calculated and are listed in Table 1.

The half vector dispersion $\delta\phi$ was corrected for the 
dispersion due to observational error \citep[see:][for discussion]{cnwk04}.
B$_{pos}$ was calculated from these
parameters by the CF method. The ratio of the observed mass-to-flux
ratio to the critical mass-to-flux ratio, $\lambda$, was calculated from
the mass and magnetic field strength, and corrected by a mean geometrical
factor of 3, following \citet{cnwk04}.

\begin{table}
\begin{center}
\begin{tabular}{lcc}
\hline
Source & L1498 & L1517B \\
\hline
Distance (pc) & 140 & 140 \\
Diameter (pc) 	         & 0.04 $\times$ 0.04 & 0.03 $\times$ 0.02   \\
$M_{obs}$ (M$_{\sun}$) 	 & 0.3 $\pm$ 0.1 & 0.3 $\pm$ 0.1 \\
$\overline{N}(H_{2})$ (10$^{22}$ cm$^{-2}$) & 1 $\pm$ 0.5     & 3 $\pm$ 2 \\
$\overline{n}(H_{2})$ (10$^{5}$ cm$^{-3}$ ) & 1 $\pm$ 0.5 & 6 $\pm$ 3 \\
$\delta\phi$ ($^\circ$) 	    & 40 $\pm$ 7    & 30 $\pm$ 5    \\
$B_{pos}$ ($\mu$G)		 & 10 $\pm$ 7   & 30 $\pm$ 10   \\
$\lambda_{obs}$			    & 8 $\pm$ 5    & 7 $\pm$ 4	 \\
$\lambda_{corr}$		    & 3 $\pm$ 2     & 2 $\pm$ 1   \\
$M_{B,crit}$ (M$_{\sun}$)	  & 0.1 $\pm$ 0.07 & 0.15 $\pm$ 0.1\\
$\Delta V$ (kms$^{-1}$)		  & 0.204 $\pm$ 0.01 & 0.195 $\pm$ 0.01 \\
$M_{\Delta V}$ (M$_{\sun}$)  	 & 0.8 $\pm$ 0.13 & 0.4 $\pm$ 0.15 \\
$M_{B,vir}$ (M$_{\sun}$)  	& 0.9 $\pm$ 0.2 & 0.55 $\pm$ 0.25 \\
$M_{P,obs}$ (M$_{\sun}$)  	& 0.8 $\pm$ 0.4  & 0.6 $\pm$ 0.3 \\
\hline
 \end{tabular}
 \end{center}
\caption{Calculated parameters of the prestellar cores L1498 and L1517B.
For the distances we followed KWA. The sizes were measured from Figures 1 
\& 2. The masses M$_{obs}$ and column densities $\overline{N}(H_{2})$
were estimated from the submillimetre photometric
mapping of KWA. The volume densities $\overline{n}(H_{2})$
were calculated assuming that the
line-of-sight dimension of each core was the mean of its other two
dimensions. The half vector dispersion $\delta\phi$ was corrected for the 
dispersion due to observational error. The plane-of-sky B-field strength
B$_{pos}$ was calculated from these
parameters by the CF method. The ratio of the observed mass-to-flux
ratio to the critical mass-to-flux ratio, $\lambda_{obs}$, was calculated from
the mass and magnetic field strength, and corrected by a mean geometrical
factor of 3 to produce the corrected value $\lambda_{corr}$.
The critical mass that could be supported by the field $M_{B,crit}$ 
was calculated from this. 
The linewidths $\Delta V$ are from \citet{taff04}.
The (thermal plus nonthermal) virial mass $M_{\Delta V}$
was calculated from the linewidth.
The total virial mass $M_{B,vir}$ was calculated by adding together
$M_{B,crit}$ and $M_{\Delta V}$. In both cases $M_{B,vir}$ is larger than
$M_{obs}$, although $M_{obs}$ must be corrected for external pressure to 
obtain $M_{P,obs}$, which is seen to be consistent with $M_{B,vir}$
in both cases (see text for discussion).}
\end{table}
\label{tab1}

For L1498 we calculated the size, mass and density of the core peak only.
To do this
we measured the flux density in the northern circular aperture on Figure~1
using the KWA data. We corrected for the chop
function along the long axis of the core, as discussed in section 3.1 above.
We used this to calculate the column density and hence the mass.
Similarly we only measured the half vector dispersion 
of the thirteen half vectors
on the core peak, so as to exclude from our calculations
the large turn of the apparent field direction from the
core peak to the filamentary tail discussed in section 3.1 above.

Our calculations using the CF method give us a plane-of-sky magnetic
field strength for L1498 of $\sim$10 $\pm$ 7 $\mu$G. This yields a 
geometrically-corrected ratio, $\lambda_{corr}$, of the observed
to critical mass-to-flux ratios of 3 $\pm$ 2.
Zeeman measurements of the line-of-sight magnetic field strength in
L1498 \citep{levin2001} were taken at the central position of
the core, marked by a star on Figure~1. These gave a field
strength of 48 $\pm$ 31 $\mu$G in the line of sight. These two
numbers are comparable, to within errors, so we are not seeing
the field at any particularly special angle. Adding the two components
in quadrature gives a total field strength of 49 $\pm$ 33 $\mu$G.
Hence we see that our geometrical correction factor of $\sim$3 is
consistent to within the errors, and we therefore leave it unchanged
to preserve comparability between the results for different cores.
From our derived value of $\lambda_{corr}$ we calculate the magnetic
critical mass to be 0.1 $\pm$ 0.07~M$_\odot$.

For L1517B we calculated the half vector dispersion separately in each of 
the three ovals shown in Figure~2 above, and then took the weighted
mean of the three dispersions. Thus we excluded any contribution to
the dispersion from the large-scale twist in the magnetic field discussed
in section 3.2 above. Hence the CF method yields a plane-of-sky magnetic
field strength for L1517B of $\sim$30 $\pm$ 10 $\mu$G. This gives a 
geometrically-corrected ratio, $\lambda_{corr}$, of the observed
to critical mass-to-flux ratios of 2 $\pm$ 1.
We note that in both cores the half vector dispersion is greater than the
recommended upper limit of 25$^\circ$ for finding accurate CF results
(Ostriker, Gammie \& Stone, 2001), 
so our values may be somewhat more uncertain as a result of this.
We obtain a magnetic critical mass for L1517B of
0.15 $\pm$ 0.1~M$_\odot$.

\citet{taff04} observed L1498 and L1517B in multiple transitions of a 
range of nitrogen- and carbon-based molecules. They recorded respective 
mean intrinsic N$_{2}$H$^+$ FWHM linewidths of 0.204 and 0.195 kms$^{-1}$
respectively, as listed in Table 1. 
We note that these authors found the same linewidths for NH$_3$.
They also found the non-thermal linewidths $\Delta V_{NT}$ to be
0.12 and 0.11 kms$^{-1}$ respectively, as used in the CF calculation above.
These non-thermal linewidths are approximately a third to a half of 
those found for the three cores studied by \citet{cnwk04}.

We used the total linewidths to calculate the (thermal plus non-thermal)
virial masses of the cores, M$_{\Delta V}$,
that could be supported virially by these linewidths, 
and found them to be 0.8 and 0.4 M$_\odot$ for L1498 and L1517B 
respectively. We list these in Table 1. 
We note that the mass able to be supported in this way is greater than the
observed mass in each case.

The parameters listed in Table~1 show that L1498 and L1517B are rather
similar in terms of their properties such as size, mass, density and 
velocity dispersion. Therefore, the fact that their field
morphologies are very different from one another, as
discussed in section 3 above, is somewhat surprising.
Table 1 shows that both cores appear to exceed their critical
mass-to-flux ratios by a factor of $\sim$2--3. We also see that the
(thermal plus non-thermal) virial mass, M$_{\Delta V}$,
in each case is larger than the observed mass. 

We can estimate the total virial mass, M$_{B, vir}$, 
(excluding the effects of external pressure)
by adding together the (thermal plus non-thermal)
virial mass, M$_{\Delta V}$,
and the magnetic critical mass, M$_{B, crit}$.
We find values of 0.9 and 0.55 M$_\odot$ respectively.
We list these in Table 1.
In both cases we see that the total virial
mass, M$_{B, vir}$,
is larger than the observed mass of the core.

In the foregoing we have simply calculated the masses in the regions
for which we have measured the magnetic field strengths. However, we
know from KWA that the total mass of each core is greater than simply
the mass of the central region. Hence we account for the extra mass
by considering this as an external pressure acting on the central region
in each case.

We can estimate this external pressure 
for a spherically symmetric core using
the expression $4\pi R^{3} n_{ext} kT$,
where k is Boltzmanns's constant, T is the core temperature 
(found from KWA), $n_{ext}$ is the external
surface density and R is the radius under consideration.
$n_{ext}$ was calculated by measuring the difference in mass between
apertures of radii $R$ and $R+dr$ and assuming that this mass came 
from a shell of material of thickness $dr$.
The density in this shell was found to be $\sim 1 \times 10^{5}$ cm$^{-3}$
for L1498 and $\sim 3 \times 10^{5}$ cm$^{-3}$ for L1517B. The value
for L1498 is in fact consistent with a uniform density out to the 
maximum radius probed.

We can incorporate this, as a correction for external pressure, to our
observed mass M$_{obs}$, to obtain M$_{P, obs}$
(c.f. Ward-Thompson 2002; Ward-Thompson et al. 2006).
We find that the additional pressure translates into
an equivalent additional mass of
0.5 and 0.3 M$_\odot$ in L1498 and L1517B respectively. 
Adding these to M$_{obs}$ we obtain values of M$_{P, obs}$
of 0.8 and 0.6 M$_\odot$ in L1498 and L1517B respectively.
These values are listed in Table 1. We estimate that the
error-bars on these values could be as high as $\sim$ 50\%,
nevertheless we see that M$_{B, vir}$ $\sim$ M$_{P,obs}$ for
both cores, and thus both are consistent with being virialised.

These two cores appear somewhat different from the three cores studied by
Crutcher et al. (2004), in that they have signicantly lower non-thermal
linewidths and magnetic field strengths. We commented above that the
three cores previously studied (L1544, L183 \& L43) were all in the 
`bright' core category of KWA, whereas L1498 and L1517B studied here both
lie near the border-line between the `bright' and `intermediate' cores
of KWA. Therefore we postulate that perhaps we are seeing a trend for
the denser `bright' cores to have higher non-thermal linewidths and
magnetic field strengths. In addition, for the denser cores thermal and 
non-thermal (including magnetic) support are approximately equal, while 
for the lower density cores, thermal support dominates.

\section{Conclusions}\label{conclusion}

We have mapped linearly polarized dust emission from the prestellar cores 
L1498 and L1517B at a wavelength of 850~$\mu$m and used the measurements to
determine the plane-of-sky magnetic field orientation in the cores.
In L1498 we saw a magnetic field on the core peak that is
offset from the core short axis by $\sim$19$^\circ$ $\pm$ 12$^\circ$. 
To the southeast of the peak, in the filamentary tail, we saw that the 
magnetic field has rotated to lie almost parallel to the long axis of the
filament. We hypothesised that the field in the core may have
decoupled from the field in the filament that connects the core
to the rest of the cloud. We used the CF method to measure the
magnetic field strength in the core of L1498 and found it
to be $\sim$10 $\pm$ 7 $\mu$G.

In L1517B we saw a more gradual turn in the field direction 
from the northern part of the core to 
the south. This appears to follow a twist in the filament in
which the core is buried, with the field staying at a roughly 
constant offset of $\sim$25$^\circ$ $\pm$ 6$^\circ$ to the short axis 
of the filament. The CF method was used to calculate the magnetic field
strength in L1517B, which was found
to be $\sim$30 $\pm$ 10 $\mu$G. Both cores were seen to 
be roughly virialised.

\section*{Acknowledgments}

The James Clerk Maxwell Telescope is operated by the Joint Astronomy 
Centre on behalf of the Particle Physics and Astronomy Research Council 
of the United Kingdom, the Netherlands Organisation for Scientific 
Research, and the National Research Council of Canada.
SCUBA and SCUBAPOL were built at the Royal Observatory, Edinburgh. 
The observations were carried out during observing runs
with reference numbers M99BU38 and M04AU26.
JMK acknowledges PPARC post-doctoral support
at Cardiff University, as well as NSF post-doctoral 
support at the University of Illinois under grant NSF AST 02-28953,
that enabled him to work on this project.
DWT was on
sabbatical at the Observatoire de Bordeaux whilst carrying out this work and 
gratefully acknowledges the hospitality accorded to him there.
RMC received partial support from grant NSF AST 02-05810.

\end{document}